\documentclass[twocolumn]{aastex7}
\usepackage{amsmath} 
\usepackage{xcolor}
\usepackage[T1]{fontenc}
\usepackage{natbib}

\newcommand\MESA{\texttt{MESA}}

\definecolor{jgcolor}{HTML}{3416B2}

\defcitealias{Krol_2026}{K26}

\begin{document}
\linenumbers
\title{In-Situ Star Formation in a Thick Disk: A Model for Metal Enrichment of Mrk 573}
\author[0009-0001-3869-5536]{Hannalore J. Gerling-Dunsmore}
\affiliation{JILA, University of Colorado and National Institute of Standards and Technology, 440 UCB, Boulder, CO 80309-0440, USA}
\affiliation{Department of Astrophysical and Planetary Sciences, 391 UCB, University of Colorado, Boulder, CO 80309-0391, USA}
\email[show]{gerlingd@colorado.edu}  

\author[0000-0003-0936-8488]{Mitchell C. Begelman}
\affiliation{JILA, University of Colorado and National Institute of Standards and Technology, 440 UCB, Boulder, CO 80309-0440, USA}
\affiliation{Department of Astrophysical and Planetary Sciences, 391 UCB, University of Colorado, Boulder, CO 80309-0391, USA}
\email[]{-}

\author[0000-0003-1012-3031]{Jared~A.~Goldberg}
\altaffiliation{NASA Hubble Fellow}
\affiliation{Columbia Astrophysics Laboratory, Columbia University, NY, NY, USA}
\affiliation{Department of Physics and Astronomy, Michigan State University, East Lansing, MI 48824, USA}
\affiliation{Center for Computational Astrophysics, Flatiron Institute, 162 5th Avenue, New York, NY 10010, USA}
\email{goldstar@columnbia.edu}

\author[0000-0003-2012-5217]{Taeho Ryu}
\affiliation{JILA, University of Colorado and National Institute of Standards and Technology, 440 UCB, Boulder, CO 80309-0440, USA}
\affiliation{Department of Astrophysical and Planetary Sciences, 391 UCB, University of Colorado, Boulder, CO 80309-0391, USA}
\email[]{taeho.ryu@colorado.edu}

\author[0000-0002-3626-5831]{Dominika \L{}. Kr\'ol}
\affiliation{Harvard-Smithsonian Center for Astrophysics, 60 Garden Street, Cambridge, MA 02138, USA}
\affiliation{Astronomical Observatory of the Jagiellonian University, Orla 171, 30-244 Kraków, Poland}
\email[]{dominika.lucja.krol@gmail.com}

\begin{abstract}
Recent observations of the nearby Seyfert galaxy Mrk 573 found that the galactic nucleus and bicone regions surrounding the jet are significantly metal-enriched. In this paper, we develop a simple model for the observed metal enrichment, assuming a population of stars in the $10 - 30 \, M_{\odot}$ range formed in-situ in the accretion disk. We assume the stars form according to a Salpeter initial mass function and generate the typical metal yields expected of field stars of the same mass. We find that in the thick disk ($H/r \sim 0.1$) case, a Salpeter population of stars can easily explain the observed metal enrichment and provide limits on the formation model; we also argue that a thin disk ($H/r \lesssim 0.01$) is incompatible with the observed metal enrichment. In our model, AGN metallicity scales most strongly with disk thickness, implying a potentially powerful observable diagnostic.  
\end{abstract}

\keywords{Accretion (14) --- Active galactic nuclei (16) --- Gravitational instability (668)  --- Magnetohydrodynamics (1964) --- Quasars (1319) ---  Supermassive black holes (1663)}

\section{Introduction}\label{sec:intro}
The Universe hosts a variety of systems featuring self-gravitating gaseous disks, e.g., protostellar and protoplanetary disks. The accretion disks of supermassive black holes (SMBHs) in active galactic nuclei (AGN) are also expected to develop substantial self-gravity in their outer regions \citep{Paczynski1978,ShlosmanBegelman1987}. However, unlike protostellar and protoplanetary disks, whose resultant objects can be directly detected, direct detection of objects forming from AGN disks has proven elusive. One of the more promising theoretical predictions of observable signatures from in-situ objects in AGN disks is metal enrichment of the AGN disk due to embedded stars. Observations of AGN with metal-enriched broad line regions (BLRs) have been proposed as potential evidence for in-situ nucleosynthesis in AGN disks \citep[e.g.,][]{HamannFerland,Juarez09}. Further, \cite{Du2014} demonstrated a correlation between BLR and narrow line region (NLR: $\sim$ kpc-scale) metallicities, strengthening the case for metal enrichment and transport within the disk. However, the poorly understood nature of the BLR has stifled attempts to answer where the metals were likely produced. 

Despite the lack of observational constraints, the potential for in-situ star formation in AGN disks has received considerable attention \citep[e.g.,][]{ShlosmanBegelman1989, Goodman2003, SG03, GoodmanTan2004, Levin2007, Chan_2024, Fryer25}, much of it focusing on how AGN embedded stars may differ from field stars found throughout the galaxy, both in individual properties, like mass and lifetime \citep[e.g.,][]{Cantiello_2021}, and in population properties, like initial mass function (IMF) \citep[e.g.,][]{BR08}. Recent work has demonstrated that the magnetorotational instability can generate a sufficiently strong toroidal magnetic field to prevent catastrophic fragmentation (i.e., cessation of a sustained flow of gas) of the disk while still permitting the formation of self-bound objects in the disk midplane \citep{GerlingDunsmore_2026}, indicating that AGN disks can survive some level of gravitational instability (GI) while forming embedded objects. While strong toroidal fields have been shown to be able to stabilize AGN disks against GI \citep[e.g.,][]{Hopkins2024FORGEdb, GerlingDunsmore_2026}, it is possible that there exist other yet undetermined mechanisms for AGN disk stabilization. However, \cite{Chen_2023} demonstrated that radiation is unlikely to contribute significantly to disk stabilization. Further, due the approximately isothermal nature of the outer regions of AGN disks (arising from the strong irradiation the region receives, as well as the low-to-moderate optical depth), sustained gravitoturbulence --- e.g., as envisaged by \cite{Paczynski1978} and \cite{Gammie2001} --- is unlikely to develop in a stabilized disk. Consequently, any mechanisms for disk stabilization that are not related to the toroidal magnetic field must still be able to operate in an environment that is globally well-approximated as isothermal (though potentially with locally non-isothermal regions); it is also possible that such mechanisms prevent the formation of bound objects. As the results of \cite{GerlingDunsmore_2026} demonstrate that MRI-driven stabilization does not preclude the formation of bound objects, disk properties congruous with magnetically dominated, MRI-driven disks are of particular interest when discussing in-situ object formation in AGN disks. 

It is commonly assumed that in-situ AGN-embedded stellar populations have a top-heavy IMF, and that individual stars may have longer lifetimes and produce higher metal yields. With sufficiently large populations of stars, or sufficiently high metallicity stars, metal enrichment of the host AGN disk (and potentially outflows, e.g., the jet and winds) is a widely expected consequence \citep[e.g.,][]{Fan_2023, Huang23}. However, thus far, there are no conclusive observational constraints on embedded stellar properties and populations (nor self-consistent theoretical models), let alone constraints on metal enrichment of the disk. 

\cite{Krol_2026} (hereafter \citetalias{Krol_2026}) presented observations of the nearby Seyfert 2 galaxy Mrk 573, finding that the nucleus and bicones display significant metallicity enhancement. The high spatial resolution of these observations ($\sim$20 pc) allowed \citetalias{Krol_2026} to associate the observed metal enrichment with the radio lobes and soft X-ray emission, demonstrating a link between the metal enrichment and AGN activity. In particular, they found consistent metal enrichment distributed along the radio jet/lobe emission and soft X-ray emission associated with shock fronts. 

In this paper, we argue that a population of embedded massive stars formed in-situ in the AGN disk can explain the enrichment of Mrk 573 reported in \citetalias{Krol_2026}, without requiring any exotic stellar properties.  We assume an SMBH mass of $M_{\bullet} = 2 \times 10^{7} M_{\odot}$ and a mass accretion rate of $\dot M_{\bullet} = 0.75 \dot M_{\rm edd} \approx 0.3 M_{\odot} \rm yr^{-1}$ \citep{revalski18}. 
In Sec.~\ref{sec:metallicity}, we discuss the observed metallicity of Mrk 573 and present our enrichment model. Metal ejection (and retention) in our model is discussed Sec.~\ref{sec:ejecta}. We discuss some implications of our model in Sec.~\ref{sec:diskcussion}, and conclude in Sec.~\ref{sec:end}.

\begin{figure}
\includegraphics[width=\columnwidth]{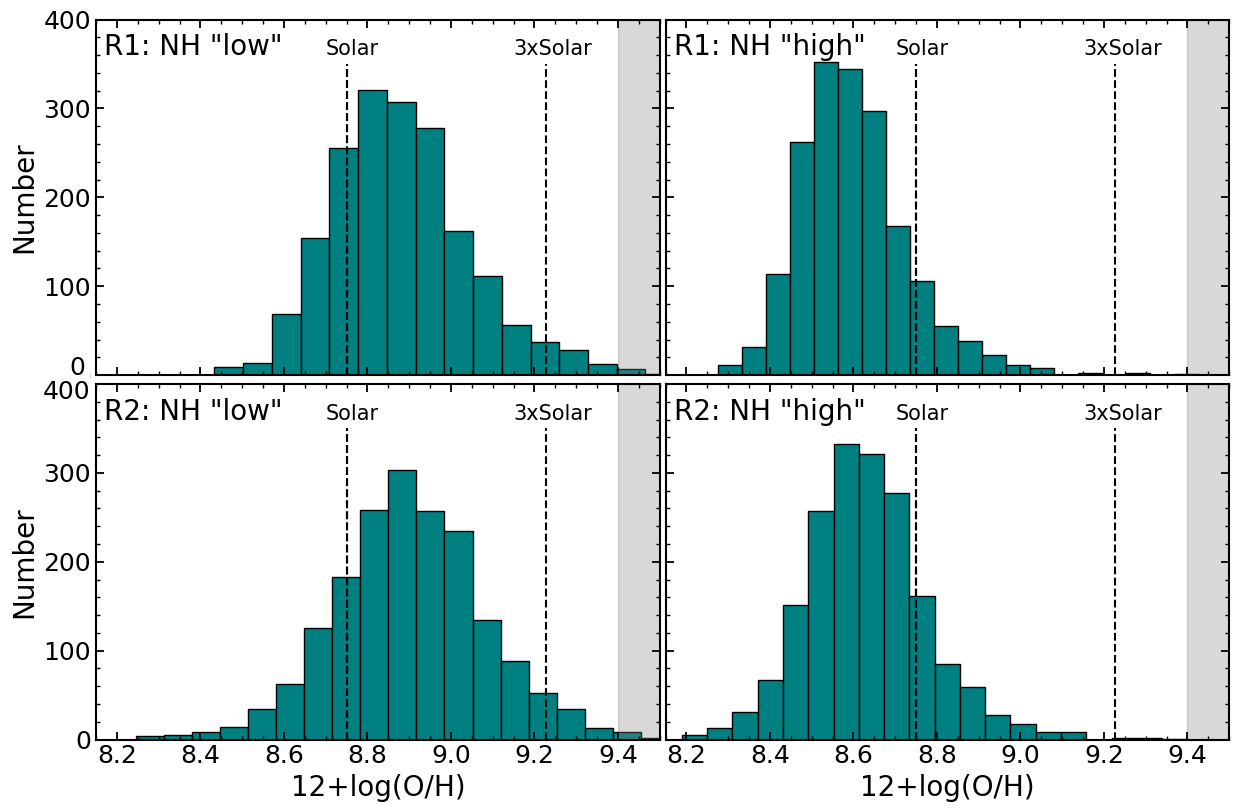}
\caption{Histograms of pixels in the \citetalias{Krol_2026} observations of Mrk 573, binned by inferred metallicity from scaling relations for N/O--O/H. The top row of histograms shows the R1 diagnostic, while the bottom row shows the R2 diagnostic; the left column shows the metallicities if assuming a low scaling relation for N/O--O/H, while the right column corresponds to a high scaling relation. Explanation of the R1 and R2 diagnostics, as well as the details of the scaling relation assumptions, can be found in Appendix~\ref{app:obs}.}
\label{fig:metalmass}
\end{figure}

\section{Enrichment Model for Mrk 573} \label{sec:metallicity}
The provenance of metals in metal-enriched AGN has been challenging to determine, due to both the theoretical challenges discussed in Sec.~\ref{sec:intro} and the historical limitations on observational spatial and spectral resolution. Lack of detectable star formation in high spatial resolution observations and the spatial association of enriched regions with radio and X-ray emission heavily disfavor ex-situ nucleosynthesis as a realistic scenario for the super-solar metallicity of Mrk 573.  Here, we briefly discuss the metal mass measured by \citetalias{Krol_2026}, and then present a model for in-situ stellar nucleosynthesis and subsequent enrichment of the AGN.  

\subsection{Observed Metallicity}\label{subsec:metalobs}
The total enriched gas mass from the observations of \citetalias{Krol_2026} is $(3-5)\times 10^{5} \,M_{\odot}$. From this total enriched gas mass, we can find an inferred total metal mass, based on scaling relations for N/O--O/H (Fig.~\ref{fig:metalmass}). The metallicity, represented as $12+\log\textrm{O/H}$ \citep{Maiolino_19} was calculated using two different diagnostics (R1 and R2) and two different N/O--O/H scaling relations: ``low'' and ``high'', corresponding to galaxies' different star formation histories (see Appendix~\ref{app:obs} for further details). In Fig.~\ref{fig:metalmass}, we see that assuming a low scaling relation implies that the vast majority of pixels have super-solar metallicity. Assuming a high scaling relation results in drastically fewer super-solar metallicity pixels, though still a significant fraction of the total pixels. Using these derived metallicities and the corresponding scaling relations we thus find that the total mass of oxygen and nitrogen for Mrk 573 is $(4-5)\times 10^{3} \,M_{\odot}$. 

\subsection{IMF, Star Formation Efficiency, and Metal Yield} \label{subsec:IMFmodel}
Models for in-situ star formation in AGN disks usually adopt a top-heavy initial mass function \citep[IMF, e.g.,][]{GoodmanTan2004, Levin2007, BR08, 10.1093/mnras/stad749} and often assume that embedded stars generate more metals than field stars of comparable mass. However, there is a lack of observational evidence for these models. Further, challenges in developing self-consistent accretion disk models (especially of the self-gravitating region) prevent the derivation of an in-situ AGN-embedded IMF consistent with disk physics. In the absence of observational evidence or well-justified theoretical models, we conservatively adopt the Salpeter IMF typically employed for field stars. The total stellar mass in a Salpeter population of stars is calculated by integrating the IMF:
\begin{equation}\label{eq:salpetermass}
  M^{\rm{tot}}_{\rm{*}} = \int^{100 M_{\odot}}_{0.1 M_{\odot}} k M^{-2.35} M dM
\end{equation}
where the bounds of $0.1 M_{\odot}$ and $100 M_{\odot}$ are typically employed for field star populations and $k$ is the normalization factor. We assume that stars in the mass range $10 - 30 M_{\odot}$ contribute to enrichment. Stars below $10M_{\odot}$ produce so few metals that their contributions would be negligible. Stars more massive than $30 M_{\odot}$ are unlikely to explode after core collapse and instead, are expected to collapse directly to black holes, with little of their generated metals ejected into the environment \citep[see, e.g.][]{Sukhbold2016}. 

As our model considers a population of stars in the range $10 - 30 M_{\odot}$, enrichment must take place over a period of at least $\sim$24 Myr (the lifetime of a $10 M_{\odot}$ star). As a fiducial case, we will assume an enrichment period of $50$ Myr, to allow two full $10 \, M_{\odot}$ lifetimes. This is well within the expected range for AGN duty cycles: the active phase of an AGN is expected to last $10 - 10^{3}$ Myr \citep{Martini_2001, Marconi04}, though it may be broken up into several smaller periods within that time frame \citep{Schawinski15, CB24}. We will also assume that the accretion rate is constant at $\approx 0.3 \, M_\odot$ yr$^{-1}$.  These assumptions imply that a total gas mass of $1.5 \times 10^{7}\,M_{\odot}$ moves through the disk during the enrichment period.

\begin{deluxetable*}{lllll}
    \tablewidth{0pt}
    \tablecaption{Stellar properties used in our model. $M_{*}$ corresponds to stellar mass; the mass range in parentheses is used to calculate the total number of stars within that range, which are conservatively assumed to have the same lifetime and generate the same metal mass as $M_{*}$. $N_{*}$ is the number of stars within a mass range, assuming star formation efficiency parameter $\eta_{*} = 0.2$. $\tau_{*}$ is stellar lifetime, $M^{\rm tot}_{\rm Z}$ is the total metal mass generated over the star's lifetime, and $M^{\rm ej}_{\rm Z}$ is the metal mass ejected into the disk after the star goes supernova (with the rest of the metals being bound in the resultant neutron star). \label{tab:starprops} 
    }
    \tablehead{
    \colhead{$M_{*}$ [$M_{\odot}$]} & \colhead{$N_{*}$} & \colhead{$\tau_{*} [\rm Myr]$} & \colhead{$M^{\rm tot}_{\rm Z} [M_{\odot}]$} & \colhead{$M^{\rm ej}_{\rm Z} [M_{\odot}]$}
    }
\startdata
  10 (10 - 12)  & 3700 &  24 & 2.5 & 1.0 \\
  12 (12 - 15)  & 3500 &  18 & 3.0 & 1.5 \\
  15 (15 - 18)  & 2100 &  13 & 5.0 & 3.0 \\
  18 (18 - 20)  & 1000 &  11  & 6.0 & 4.0 \\
  20 (20 - 25)  & 1700 &  9.7 & 7.0 & 5.0 \\
  25 (25 - 30)  & 1100 & 7.7  & 9.5 & 7.5 \\
\enddata
\end{deluxetable*}

For field stars formed from molecular clouds, the star formation efficiency is believed to be a few percent, i.e., a few percent of the total gas mass will form stars, while the rest of the gas is blown away by stellar feedback. In an AGN disk, this picture is unlikely to hold. The disk is continually re-supplied with gas captured from the environment and experiences significant Coriolis force, which acts to cohere the flow and would tend to stabilize the disk against bursts of feedback. We thus make no assumption about the star formation efficiency of the disk in our model at the outset, and instead use $\eta_{*}$ to parametrize star formation efficiency within the disk. The total stellar mass formed over the enrichment period  is $M^{\rm tot}_{*} = 1.5 \times 10^{7}\eta_{*} \, M_{\odot}$. Plugging this into Eq.~\ref{eq:salpetermass}, we find $k =  2.6 \times 10^{6}\eta_{*} M^{1.35}_{\odot}$. 

In order to estimate reasonable metal yields for the population of efficient enrichers, we performed a suite of 1D stellar evolution calculations using the open-source stellar evolution software Modules for Experiments in Stellar Astrophysics (\texttt{MESA}; \citealt{Paxton2011,Paxton2013,Paxton2015,Paxton2018,Paxton2019,Jermyn2023}) revision-24.08.1. We model single stars with zero-age main-sequence (ZAMS) masses listed in Table~\ref{tab:starprops}, following closely the setup described in Sec.~7.2 of \citet{Jermyn2023}. We adopt modest step convective overshooting calibrated to observations \citep[adopting the][values]{Brott2011a,Brott2011b} and a mixing length parameter $\alpha=2.5$ \citep[see][]{Chun2018,Goldberg2022a}. We present non-rotating stellar models to provide a conservative estimate on the yields. A similar setup has been employed in the context of stripped stars \citep{DornWallenstein2025,TDWZenodo}. Further detail on the \texttt{MESA} setup can be found in Appendix~\ref{app:mesastars}, and a full reproducibility package is available \href{https://github.com/aurimontem/MESA_massive_star_setup#}{here}\footnote{github link to be replaced with Zenodo repository upon manuscript's acceptance}.

We adopt an initial metallicity at ZAMS of $Z_{\rm{ZAMS}} = 0.1 Z_{\odot}$ for our fiducial simulations. We also calculated models at an initial metallicity of $Z_{\rm{ZAMS}} = Z_{\odot}$, and found that initial metallicity plays a weak role in metal yield and stellar lifetime (at the $\sim10\%$ level). We likewise varied the strength of stellar winds, considering strong (3.3$\times$ fiducial), moderate (fiducial), and weak (0.33$\times$ fiducial) wind efficiency, which also had little impact on the metal yield per stellar model ($\lesssim5\%$). 

Our estimated metal yields for 6 mass groups are shown in Table~\ref{tab:starprops}. We integrate: 
\begin{equation}\label{eq:salpeterN}
  N^{M_{\rm upper}}_{M_{\rm lower}} = 2.6 \times 10^{6}  \eta_{*}\int^{M_{\rm upper}}_{M_{\rm lower}} M^{-2.35} dM
\end{equation}
to get the total number of stars within the mass range $M_{\rm lower} \leq M_{*} \leq M_{\rm upper}$. To obtain a conservative estimate of the metal yield, we assume that all stars in a mass range produce the same metal yield as the lower bounding mass. We thus find the total metal mass produced during the enrichment period by multiplying $N^{M_{\rm upper}}_{M_{\rm lower}}$ by the metal yield of a star of $M_{\rm lower}$ (see Table~\ref{tab:starprops} for further explanation). Summing all metal yields for the considered mass ranges gives a total metal mass of $M^{\rm tot}_{\rm Z} =  2.0 \times 10^{5} \eta_{*} \, M_{\odot}$. If $\eta_{*} < 0.02$, the disk does not produce enough metals to match the observations of \citetalias{Krol_2026} even if the entire metal yield contributes to the outflow. More realistically, most of the disk gas, including the metals, will be swallowed by the black hole.  If only $\sim 10 \%$ of the integrated disk mass is loaded into the biconal region, then the star formation efficiency must be rather large, $\eta_* \gtrsim 0.2$. The total metal yield for each stellar mass range can also be found in Table~\ref{tab:starprops}, under $M_{\rm Z}^{\rm tot}$.  

While massive field stars usually form in binaries \citep[e.g.,][]{Sana+2012,Sana+2013,deMink2013,Offner2023}, we will leave their inclusion to future, more detailed, work. Here, we aim to provide conservative limits on the disk enrichment possible from a population of in-situ stars.  
Nucleosynthesis in a binary system in which both stars are within our population of efficient enrichers is expected to be minimally impacted (or enhanced; see, e.g., Fig.~5 of \citealt{JzMa2025}) as compared to the two objects evolving separately.
Additionally, once the stars reach the onset of core helium burning, the stellar envelope expands dramatically and functionally de-couples from the stellar core, at least in terms of total metal yield.\footnote{In general, stripping of the H-rich envelope during the crossing of the Hertzprung Gap causes \textit{increased} total metal yield at fixed He core mass; see e.g. \citet{Laplace2021}. Similarly, massive stars accreting on the main-sequence tend to show rejuvenation and thereby core evolution comparable to \textit{more} massive stars \citep[e.g.][]{Renzo2023}. If the He core is already formed at the time of accretion, an accretor's core and thereby metal mass remain comparable to that of a single star \citep[see, e.g.][]{Farrell2020}.} Thus, after the He core has developed, the envelope can become stripped without substantially impacting the total metal yield at the time of explosion. We further address the adequacy of single-star calculations for this work in Appendix~\ref{app:mesastars}. 

\subsection{Disk Properties} \label{subsec:diskmodel}
We next consider the star-forming region of the disk.  Star formation can only occur in a gravitationally unstable gaseous region. In a disk, the strength of self-gravity is parametrized using the Toomre stability parameter \citep{Toomre1964}:
\begin{equation}\label{eq:QT}
  Q_{\rm{T}} = \frac{c_{\rm{s}} \Omega}{\pi G \Sigma}
\end{equation}
where $c_{\rm{s}}$ is the local sound speed, $\Omega = \sqrt{G M/r^{3}}$ is the orbital frequency at cylindrical radius $r$ from a central object of mass $M$, and $\Sigma$ is the surface density of the region. 

In typical AGN disks, $Q_{\rm T}$ is expected to decrease outward. When $Q_{\rm{T}}$ is large (e.g., in the inner regions of the accretion disk), the disk is strongly stable against gravitational instability (GI); when $Q_{\rm{T}} < 1$ (i.e., self-gravity dominates the internal pressure), runaway gravitational instability develops, leading to disk fragmentation. The radius at which $Q_{\rm T} \simeq 1$ determines the outer edge of the stable, sustained accretion disk. While only the gas sound speed is considered in the conventional definition of $Q_{\rm T}$, additional contributions to  pressure support should be included \citep[e.g.,][]{KimOstriker01,Lizano_2010} to more accurately characterize GI \citep[e.g., magnetic fields;][]{GerlingDunsmore_2026}. Therefore, we introduce $c_{\rm{i}}$, the characteristic speed of the physical process(es) that generate the leading order contribution(s) to the pressure within a disk, and replace $c_{\rm s}$ by $c_{\rm i}$ in Eq.~\ref{eq:QT}. 
For a gas-dominated disk, $c_{\rm{i}}$ is the sound speed; in a magnetically dominated disk, $c_{\rm{i}}$ is the Alfvén speed. The disk scale height, $H$, can be written as $H = \frac{c_{\rm{i}}}{\Omega}$, and we can re-write the Toomre stability parameter as
\begin{equation}\label{eq:QTfinal}
  Q \approx \left( \frac{H}{r} \right) \left( \frac{M_{\bullet}}{M_{\rm{disk}} (<r)} \right),
\end{equation}
where $M_{\bullet}$ is the mass of the SMBH, $\frac{H}{r}$ is the aspect ratio of the disk, $M_{\rm disk}(<r)$ is the enclosed disk mass within radius $r$, and we drop the subscript $T$. 
Using Eq.~\ref{eq:QTfinal} and taking the radius at which $Q = 1$ to be the outer radius of the disk, we find that the total instantaneous disk mass is $M^{\rm tot}_{\rm disk} \approx \frac{H}{r} M_{\bullet}$. For a thin ($H/r = 0.01$) and thick ($H/r = 0.1$) disk, the respective total instantaneous disk masses for a $2 \times 10^{7} M_{\odot}$ SMBH are $M^{\rm inst}_{\rm disk} = 2 \times 10^{5}$ and $2 \times 10^{6} \, M_\odot$. 

Since the instantaneous disk mass is significantly lower than the total captured mass over the enrichment period ($1.5 \times 10^{7} M_{\odot}$), in both thin and thick disk cases, we can consider the disk to be in a steady state.  As the inflow time (set by the disk dynamics) is shorter than the enrichment time (set by the stellar lifetime), the metals generated by stellar nucleosynthesis can be injected into the disk at any radius, mix with the non-enriched gas in the disk, and be transported to the jet launching region sufficiently rapidly to get loaded into the bicone region along with the outflow. Unlike the gas, which flows through the disk in a steady state, the newly formed stars will accumulate in the absence of significant stellar migration, 
reaching a total mass $M_*^{\rm tot} = 2 (\eta_*/0.2) (H/0.1 r)^{-1} M^{\rm inst}_{\rm disk}$; we find this expression by re-writing our earlier expression for $M_*^{\rm tot}$ in terms of the instantaneous disk mass. Star formation is probably suppressed when $M_*$ significantly exceeds the instantaneous disk mass, since the gas becomes more attracted to the existing stars than to its self-gravity.  This suggests that, for the parameters of Mrk 573, the enrichment must occur in a relatively thick disk, with $H/r \gtrsim 0.1$.   

As an independent check on disk thickness, 
we consider the average maximum size of an in-situ object contracting under its own self-gravity as a function of disk properties. If we assume $M_{\rm{disk}} = \Sigma \pi r^2 \approx \rho H \pi r^2$, where $\rho$ is the disk's average density, we can use Eq.~\ref{eq:QTfinal} to find $\rho \approx (H/r) M_{\bullet} Q^{-1} / (H\pi r^2) \approx M_{\bullet} Q^{-1} / (\pi r^{3})$. The maximum mass of a bound object that can form embedded in an accretion disk is $M^{\rm{max}}_{\rm{bound}} \approx \rho (\frac{4}{3}\pi H^{3})\approx M_{\bullet}Q^{-1}(H/r)^3$, as the largest radius an embedded spherical object can have is the disk scale height. For our considered disk aspect ratios of $\frac{H}{r} = 0.01$ and $0.1$, and assuming $Q = 1$, this gives $M^{\rm max}_{\rm{bound}} = 27$ and $2.7 \times 10^{4}  M_{\odot}$, respectively. While the thick disk can easily form sufficiently large objects to explain a population of efficient enrichers, for the thin disk the available mass is marginal. This independently supports our deduction that the disk must be relatively thick. Further, by comparing the gas density of the disk in the star forming region ($\rho \approx 10^{-18} \, \rm g \, cm^{-3}$) to the gas density in the outer envelop of the considered stars ($\rho_{*} \approx 10^{-10} \, \rm g \, cm^{-3}$), we can see that the disk is unlikely to exert sufficient pressure on the embedded stars to increase their fusion rate. Thus, we can assume that the stars in our model undergo normal evolution while on the main sequence. As we discussed in Sec.~\ref{subsec:IMFmodel}, evolution of the stellar core is insensitive to environment after detachment from the outer envelope. Thus, we can conclude that the metal yields from the stars in our model evolve as expected for field stars of the same masses. 

Such large disk aspect ratios greatly exceed the values that can be supported by gas pressure, and are more characteristic of support by a combination of strong magnetic field and magnetically dominated turbulence \citep{begelman23,begelman24,Hopkins2024FORGEdb}.  Moreover, moderately self-gravitating, magnetically elevated disks may be more resilient against wholesale fragmentation than their thinner gas pressure-dominated counterparts \citep{GerlingDunsmore_2026}. 

These considerations allow us to estimate the distance from the SMBH, $r_*$, at which stars form in Mrk 573. For the adopted enrichment time $50 \, \rm Myr$ and the accretion rate of $0.3 \, \rm M_{\odot} \, yr^{-1}$, the disk inflow timescale is
\begin{equation}\label{eq:tin}
  t_{\rm in} \approx 66 \left(\frac{H}{r} \right) \ {\rm Myr},
\end{equation}
with an inflow speed of $v_{\rm in} \approx \alpha (H/r)^\nu v_{\rm K}(r_*)$, where the viscosity parameter $\alpha \approx 1$ for a magnetically dominated disk \citep{begelman23a}, $v_{\rm K}$ is the Keplerian speed ar $r_*$, and $\nu = 2$ for a disk dominated by internal viscosity and $\nu = 1$ for inflow driven by large-scale magnetic torques \citep[e.g.][]{blandford82,ferreira93}. Setting $r_* = t_{\rm in} v_{\rm in}$, we find that $r_* \sim 7-30$ pc for $H/r = 0.1$, where the greater value applies for $\nu = 1$. This corresponds to $\sim 10^7 r_{\rm g}$ for a $2 \times 10^7 \, M_\odot$ SMBH, where $r_{\rm g} = GM_{\bullet}/c^2$ is the gravitational radius.  This is well outside the BLR, which is typically located at $\sim 10^4-10^5 r_{\rm g}$. 

It is worth noting that for a $2\times10^{7} M_{\odot}$ SMBH, the sphere of influence derived from the M-$\sigma$ relation is approximately 10 pc, coinciding with our derived radius for the star forming region. We can estimate the migration time of stars from this region using $\tau_{\rm mig} = 0.5Q\frac{M_{\bullet}}{M*}\frac{H}{r}\Omega^{-1}$ \citep{Ward97}, and find a migration time of approximately $10^{9} \, \rm yr$ for our considered mass range of stars. We thus find that migration times are far longer than metal enrichment times at these distances, justifying our neglect of the migration process. 

\section{Metal Ejection \& Ejecta Trapping}\label{sec:ejecta}
All stars in our population of efficient enrichers are expected to undergo iron core-collapse, leaving behind a compact remnant. Most of the stars are expected to explode (see, e.g., \citealt[][]{Janka2012,Sukhbold2016,Couch2017,Boccioli2023,Burrows2024} and references therein), resulting in a neutron star with the rest of the stellar material ejected into the environment. Each neutron star binds $\approx1.4 - 2 M_{\odot}$ of the metals generated by the star's evolution. We assume for simplicity that the neutron star removes $\approx1.5 M_{\odot}$ of metals from our two smallest mass ranges and $\approx2 M_{\odot}$ of metals from the larger mass ranges \citep[see, e.g.][]{Boccioli2024}. The resultant metal yields in the ejecta for each mass range are shown in Table~\ref{tab:starprops}, under $M^{\rm{ej}}_{\rm{Z}}$. 

At $10^{7} r_{\rm g}$, the escape velocity from the SMBH's gravitational potential is approximately $100$ km s$^{-1}$. This is far lower than core-collapse supernova ejecta velocities ($3000 - 10000$ km s$^{-1}$), which unimpeded would escape to infinity. However, the supernova ejecta will decelerate rapidly once the entrained mass from the disk becomes comparable to the ejecta mass \citep{SNRemnantReview}. 
From this point, the ejecta speed scales as $v_{\rm ej}\propto r_{\rm ej}^{-19/6}$ while the blast wave that absorbs the supernova energy and runs ahead slows down as $v_{\rm bl}\propto r_{\rm bl}^{-3/2}$; here, $v_{\rm ej}$ and $v_{\rm bl}$ are the velocities of the ejecta and blast wave, respectively, while $r_{\rm ej}$ and $r_{\rm bl}$ are the travel distances of the ejecta and blast wave. For a disk density of $\sim 10^{-18} \rm \, g \, cm^{-3}$, even an extremal $25M_\odot$ of SN ejecta sweeps up its own mass by time it has expanded to $\sim 10^{17} \rm \, cm$. This density is calculated assuming all of the disk mass at a given radial position in the disk is contained within one scale height; in reality, the disk density is likely somewhat lower, but spread over a much larger vertical extent. What is relevant to ejecta trapping is the total mass swept up; therefore, a disk that has half the density but twice the vertical extent will trap ejecta approximately as effectively. This argument neglects more complicated thermodynamic effects of the ejecta's interaction with the disk material, which will be necessary to determine potential observable impacts on disk spectra due to embedded stellar populations. Such a study exceeds the scope of this paper.  

In a thick disk, at our derived star formation radius of $10^{7} \, r_{\rm g}$, the disk height is $3 \times 10^{18} \rm \, cm$, by which point the metal-rich ejecta will remain bound. 
We thus see that the ejecta are easily trapped by the flow of the disk long before they reach the disk surface in the thick disk case. Note that in the thick disk case, the blast wave (which mostly consists of  shocked disk material but not the metal-enriched ejecta themselves) might reach the disk surface with a velocity comparable to or larger than the local sound speed. This may disturb the disk but should not affect the trapping of the metals. Observable signatures of such breakout blastwaves may be worth studying.

This also provides a third argument for a thick disk since, in contrast, the ejecta could potentially punch through a disk with $H/r \lesssim 0.01$, as the disk height is only $3 \times 10^{17} \rm \, cm$ and stars would have to go supernova almost precisely in the middle of the disk in order to prevent their metals from escaping to infinity. In contrast, a star would have to go supernova at $z > 0.9H$ in a thick disk before the ejecta shell would break out of the disk before sweeping up enough mass.

Assuming that stars form close to the midplane, they would have to undergo significant three-body interactions in order to get kicked into an orbit that puts them at such high altitudes within the disk.  Interactions with the gas within the disk would return their orbits to the midplane over time \citep[e.g.,][]{Syer+1991}; thus, they would have to experience the dynamical kicks shortly before going supernova. Moreover, the less massive object is usually the one that gets kicked in such dynamical interactions \citep{Valtonen+2006}, indicating that the most metal-rich stars (and therefore, the stars whose ejecta shells will expand the farthest) would be unlikely to go supernova significantly off midplane. See Appendix~\ref{app:threebody} for a simple estimate suggesting that kicks from three-body encounters can safely be neglected.


\section{Implications}\label{sec:diskcussion}
\subsection{Magnetic Elevation and Disk Structure}
Our model finds that AGN disks must be fairly thick to explain significant metal enrichment; \cite{BegelmanPringle2007} found that in magnetically dominated disks, the disk thickens due to the enhanced magnetic pressure, resulting in magnetic elevation. Recent work \citep{Hopkins2024FORGEda, GerlingDunsmore_2026} found that magnetic dominance is required for an AGN disk to stabilize against GI. However, when $Q \sim \mathcal{O}(1)$, a stabilized disk can still form objects without undergoing catastrophic fragmentation resulting in total disruption of the disk. As mentioned in Sec.~\ref{sec:intro}, the strongly magnetized, strongly self-gravitating simulation presented in \cite{GerlingDunsmore_2026} developed a bound object in its midplane that persisted until the object moved out of the box. It is thus plausible that in-situ star formation in AGN disks requires magnetic elevation of the disk, and further, it is likely that as long as the disk is able to maintain magnetic dominance, the disk can continue extending radially. This would create an extended region in which $Q \sim \mathcal{O}(1)$, and therefore an extended region in which stars could form. Further study of conditions for maintaining magnetic elevation in AGN disks (and thus, the parameters of the star-forming region) is vital for developing a realistic AGN-specific IMF.

One implication of the star forming region occurring at such large radii is that all metals generated in-situ must pass through the BLR in order to get to the bicone loading region in the inner disk. This raises the possibility that the enrichment of the BLR reflects the current state of the disk's enrichment (as metal-enriched gas is transported from the star forming region inwards), while the bicone reflects the time-integrated enrichment of the disk. If the metals in the disk are predominantly formed exterior to the BLR, then at the end of the AGN cycle, the BLR will cease to have metals supplied to it long before the bicones cease to have metals injected. Modeling the temporal lag between the BLR losing metallicity and the bicones losing metallicity may allow us to determine if a given AGN is at the end of its active phase. Conversely, the same could be done for AGN that are beginning the active phase, by modeling the lag between the enrichment of the BLR and the enrichment of the bicone region. Metallicity distribution in AGN may prove a useful tool for constraining the age of a given AGN, though this diagnostic remains purely speculative at this time. 

\subsection{Comparison to Other Embedded Stellar Models} One of the major results of our model is that enriched AGN disks are likely to be thick. This holds for four reasons: maximum bound mass in a thin disk is inadequate to form sufficiently massive stars to generate the observed metal mass, the disk mass of a thin disk is too low to meet the criterion that the disk mass must exceed the total stellar mass (as otherwise, the disk ceases to function as a disk), thin disks cannot trap core-collapse supernova ejecta, and there is currently no known mechanism to permit thin disks to survive critical stability against GI (as discussed above). Most previous models of in-situ stars \citep[e.g.,][]{GoodmanTan2004, Levin2007} explicitly assume a thin disk. The thinness of the disk plays a role in one of the key conclusions of \cite{Levin2007}, which is that embedded mergers are unlikely to show significant inspiral modification from their vacuum-environment counterparts, but a gap may open in the AGN disk around a late stage merger. While determining whether these two results hold in the thick disk case exceeds the scope of this work, determining the impact of disk thickness on embedded mergers (and scenarios in which observable gaps may open in the disk) would potentially provide diagnostics of disk structure in the Laser Interferometer Space Antenna (LISA) band \citep{amaroseoane2017laserinterferometerspaceantenna}. 

Another recent model of embedded star formation and evolution is that of ``immortal stars'' \citep{Cantiello_2021, Jermyn_2022}, which posits that stars embedded in AGN disks accrete hydrogen faster than their hydrogen fuses into helium, thus prolonging hydrogen-burning as long as the stars have fresh material to accrete. Our disk model is incompatible with the conditions required for immortal stars, as the required surface density to sustain such accretion onto embedded stars would result in a disk far too massive for a $2 \times 10^{7} \, M_{\odot}$ SMBH with the radial extent we self-consistently calculate. However, the environment in our model may be sufficiently dense to increase merger rates between stars, leading to rejuvenation in the spirit of ``immortal stars.'' This could potentially result in substantially more massive stars than predicted by a simplistic IMF. 

\section{Conclusions}\label{sec:end}
In this paper, we presented a model of metal enrichment of an AGN disk by a population of embedded stars, formed in-situ. We assumed a Salpeter IMF and standard evolutionary properties of the stars, to determine whether the observations of \citetalias{Krol_2026} required the exotic stellar populations commonly proposed for AGN embedded stars. We showed that this model can explain the observed metal enrichment of Mrk 573 if the disk is thick ($H/r \sim 0.1$). We also demonstrated that a thin disk cannot explain the observed enrichment. Our main conclusions are as follows: 

\begin{itemize}
    \item The metal enrichment of Mrk 573 can be self-consistently explained using a population of stars embedded in the accretion disk, formed in-situ; a Salpeter population of stars is adequate to explain the observed enrichment, if the disk has an aspect ratio of $H/r \sim 0.1$ and $\gtrsim 20\%$ of the gas captured to the disk fragments into stars over a 60 Myr enrichment period. 
    \item Due to its thickness and ability to survive marginal stability against GI, the disk is likely magnetically elevated.
    \item The disk likely extends to approximately $10^{7} \, r_{\rm g}$, i.e., near the edge of the BH sphere of influence.
    \item The observed enrichment can possibly be explained with a slim disk ($0.01 < H/r < 0.1$), but would require very high star formation efficiency and/or an extremely top-heavy IMF. 
\end{itemize}

AGN metallicity thus shows considerable potential as an observable diagnostic for properties that are currently hard or impossible to observationally constrain. The effective use of this metric, however, requires further study of the star-forming region of AGN disks and magnetically elevated accretion disks broadly. Such investigations are essential before accurate phenomenology work can begin.

\section*{Acknowledgements}
The authors thank the referee for a helpful report, which improved the clarity of this paper. HGD acknowledges support from NASA FINESST Fellowship 80NSSC22K1753. MCB acknowledges support from NASA Astrophysics Theory Program grant 80NSSC24K0940.
J.A.G. acknowledges financial support from NASA grant 23-ATP23-0070. The Flatiron Institute is supported by the Simons Foundation.

\section*{Data Availability}
A reproducibility package for the \MESA\ setup, including all \texttt{inlists}, \texttt{run\_star\_extras} files, and run scripts, 
is available on \href{https://github.com/aurimontem/MESA_massive_star_setup/tree/main}{github} (to be replaced with a Zenodo link upon acceptance).

\bibliography{bibliography}{}
\bibliographystyle{aasjournalv7}

\appendix
\section{Mrk 573 Metal Mass Measurement} \label{app:obs}
The metallicities for Mrk 573 are inferred from the metallicity diagnostics from \cite{Zhu_2023}:
\begin{eqnarray}\label{eq1}
   R1 = R(N2,S2,{\rm H}\alpha) =\log\bigg(\frac{[N~II]\lambda6584}{[S~II]\lambda\lambda6717,31}\bigg)+  0.264\log\bigg(\frac{[N~II]\lambda6584}{{\rm H}\alpha}\bigg) 
\end{eqnarray}
and 
\begin{eqnarray}\label{eq2}
   R2 =  R(N2,S2,O3)  &&=   0.9738\log\bigg(\frac{[N~II]\lambda6584}{[S~II]\lambda\lambda6717,31}\bigg)  ,\nonumber\\
    &&+0.047\log\bigg(\frac{[N~II]\lambda6584}{{\rm H}\alpha}\bigg)  - 0.183\log\bigg(\frac{[O~III]\lambda5007}{{\rm H}\beta}\bigg).
\end{eqnarray}

Histograms in Fig.~\ref{fig:metalmass} show metallicity values from \citetalias{Krol_2026}, calculated assuming two N/O--O/H scaling relations \citep{Groves04, Dors17}: ``low'', $\log({\rm N/O}) = \log(10^{-1.732} + 10^{\log({\rm O/H})+2.19})$ (left panels), and ``high'', $\log({\rm N/O}) = 1.29\times(12+\log({\rm O/H})-11.84)$ (right panels). To calculate the masses of oxygen and nitrogen, we used the total mass of emitting gas $\sim (3-5)\times 10^5$~M$_{\odot}$ (derived based on the formula from \cite{Cano_12}, 
see K26 for details). We then calculated the masses of oxygen and nitrogen based on the derived $\log$(O/H) and  N/O--O/H scaling relations, obtaining metal masses of $\sim5\times10^3$~M$_{\odot}$ for ``low'' and $\sim4\times10^3$~M$_{\odot}$ for ``high'' N/O--O/H scalings. As is typical in discussions of AGN metallicity, we assume the total metal mass is well-approximated by the N+O mass. 

\section{Additional discussion of stellar modeling} \label{app:mesastars}

The \MESA\ Equation of State is a blend of the OPAL \citep{Rogers2002}, SCVH
\citep{Saumon1995}, FreeEOS \citep{Irwin2004}, HELM \citep{Timmes2000},
PC \citep{Potekhin2010}, and Skye \citep{Jermyn2021} equations of state.
Radiative opacities are primarily from OPAL \citep{Iglesias1993,
Iglesias1996}, with low-temperature data from \citet{Ferguson2005}
and the high-temperature, Compton-scattering dominated regime by
\citet{Poutanen2017}.  Electron conduction opacities are from
\citet{Cassisi2007} and \citet{Blouin2020}.
Nuclear reaction rates are from JINA REACLIB \citep{Cyburt2010}, NACRE \citep{Angulo1999} and
additional tabulated weak reaction rates \citet{Fuller1985, Oda1994,
Langanke2000}. We adopt the \texttt{approx21\_cr60\_plus\_co56} nuclear reaction network, which is insufficient to predict detailed core structure relevant for simulations of the core-collapse mechanism \citep{Farmer2016, Renzo2024}, but sufficient for calculating approximate overall metal yields \citep[see further discussion in][]{Farmer2016}. Screening is included via the prescription of \citet{Chugunov2007}.
Thermal neutrino loss rates are from \citet{Itoh1996}.
The adopted wind prescription follows the \citet{Brott2011a,Brott2011b} parameters adopted by \citet{Jermyn2023}, which are a blend of 
the \citet{Vink2001}, \citet{Nieuwenhuijzen1990}, and reduced (by a factor of 10) \citet{Hamann1995} wind prescriptions in their respective regimes of applicability, with an efficiency factor of 0.3 for the fiducial runs.

\subsection{Potential impacts of binary stellar evolution}
Here we briefly justify our logic in neglecting binarity in our metal yield estimates from stellar evolutionary calculations. In binaries with minimal interaction between the stars, the stellar evolution of both stars proceeds essentially unmodified from the single star case. Therefore, the metal yields from a low interaction binary would be functionally unchanged from the two stars evolving independently. In binaries where the stars merge early (while both stars are still in hydrogen or helium burning), the resultant star is significantly more massive. Metal production scales strongly with stellar mass. As illustrated in Table~\ref{tab:starprops}, the metal yield from say, two 10 $M_{\odot}$ stars is less than the metal yield from one 20 $M_{\odot}$ star. So long as both stars in the early-merger binary are under 15 $M_{\odot}$, the resultant star will most likely undergo a successful supernova after core collapse. Even if both stars are as massive as 20 $M_{\odot}$, there's still a significant probability that the resultant star will explode \citep[e.g.][]{Vartanyan2021,Gilkis2025}. Given the comparatively small number of stars over 20 $M_{\odot}$ compared to those under 15 $M_{\odot}$ (see Table~\ref{tab:starprops}), the number of early-merger binaries that result in a star that directly collapses to a black hole (and therefore not contributing to the enrichment of the disk) would be very small, and can safely be neglected.\\

In the case of binaries which undergo a late merger (after one star has begun carbon burning), the more massive star will engulf the less massive star. This will result in the more massive star producing similar metal yields as it would have produced evolving independently, and at most, prevent metal production in the less massive star. Therefore, at worst, a late merging binary will result in a reduction of a factor of two in the metal yield; while non-trivial, such a reduction does not alter the order of magnitude of the metal yields, even in the most pessimistic scenario. If the binary system develops stable mass transfer, the more massive star will usually become the donor star, which typically occurs after core hydrogen burning. As discussed in Sec.~\ref{subsec:IMFmodel}, the stellar core evolution becomes functionally ``detached'' from the envelope by core helium burning, resulting in the more massive star in the binary producing metals comparably to how it would if evolving independently. The accretion of material from the initially more massive (donor) star onto the initially less massive (accretor) star usually begins while the accretor is still undergoing hydrogen burning (and therefore the core is still ``attached'' to the envelope). Thus the accretor star may actually increase its metal yields compared to evolving as a singleton.\\

Finally, it bears mentioning that while a $<10 M_{\odot}$ star produces negligible quantities of metals, two $<10 M_{\odot}$ stars merging often results in a star massive enough to contribute non-trivial metal production. Given the vastly larger population of low mass stars compared to massive stars, inclusion of binarity in the low mass stars currently excluded in our model could entail an appreciable increase in the predicted metal yields. Thus, due to our focus strictly being metal enrichment of the disk (and subsequently, the jet), neglecting binarity's effects on our estimated metal yields from stellar evolution calculations is reasonable, if not conservative.\\

\section{Additional discussion of three-body interaction time scales} \label{app:threebody}

For a given stellar binary, the rate of three-body interactions due to a third star encountering the binary is given by: 
\begin{equation}\label{eq:nsigmav}
    \mathcal{R} = n\sigma v
\end{equation}
\noindent where $n$ is the number density of relevant stars not in a binary, $\sigma$ is the encounter cross section (typically taken to be the square of the characteristic binary separation within the system), and $v$ is the characteristic velocity of single stars relative to a given binary. As mentioned in Sec.~\ref{sec:ejecta}, the ejected/kicked object in a three-body interaction is likely to be the lowest mass object in the interaction. Thus, in order to adequately kick (or fully eject) an efficient enricher, the other two stars must either be efficient enrichers or stars so massive that their supernova will fail. The total number of potential ``kicker'' stars can then be found by integrating Eq.~\ref{eq:salpeterN} from $10 M_{\odot}$ to $100 M_{\odot}$, yielding $N_{>10} = 1.6 \times 10^{4}$.  To find $n_{>10}$, we then divide $N_{>10}$ by the volume of an annulus at $r = 10^{7} R_{\rm g}$ with a radial extent of one Jeans length \citep[$L_{\rm J} \approx 2H$;][]{GerlingDunsmore_2026}, finding $n \approx 250 \, \rm pc^{-3}$. One Jeans length is the smallest radial extent that still permits gravitational collapse into stars, thus providing an upper limit on the three-body interaction rate. The radial extent of the three-body interaction region is likely much larger, which would significantly decrease $n_{>10}$. 

Using Eq.~\ref{eq:nsigmav} with $v$ set to the Keplerian velocity of the disk at radius at which $Q \approx 1.0$ and the maximum value for $n_{>10}$, we find that for a given binary, the total three-body encounter rate over our 50 Myr enrichment period is $\mathcal{R} = 2 \times 10^{-4} a^{2} \, \rm (50 Myr)^{-1}$, where $a$ is the separation of the stars in the binary in AU. For $a = 100 \, \rm AU$, we find $2$ relevant three-body encounters per binary over the enrichment period. Only once the binary separation is taken to be at least $1000 \, \rm AU$ do we find a high encounter rate ($\approx 200$ per 50 Myr enrichment period). In field stars, binary separation is typically of order 10 AU \citep{Offner2023}; it is not currently clear what the typical binary separation would be for an in situ AGN-embedded population of stars would be. Assuming $v$ is the local Keplerian velocity, if the stellar separation is taken to be $a = 10 \, \rm AU$, we find $2 \times 10^{-2}$ three body encounters with the \textit{potential} to significantly kick an efficient enricher over the full enrichment period. 

The total three-body interaction rate is further reduced by assuming that only a small fraction of the stellar population is in binaries, and the entire stellar population forms at the same time. Moreover, in-situ stars will likely co-move with the disk and have much  smaller velocities, relative to each other, than the Keplerian velocity we have assumed. 

These calculations suggest three-body interactions are unlikely to kick a substantial number of efficient enrichers far enough off the disk midplane to impact AGN metallicity. However, future work on an AGN-specific IMF (including binary fraction and characteristic binary separation) is necessary to accurately determine the relevance of three-body interactions for in situ disk enrichment.
 
 \end{document}